\pdfoutput=1
\documentclass[american]{article}
\usepackage[T1]{fontenc}
\usepackage[latin9]{inputenc}
\usepackage[a4paper]{geometry}
\usepackage{babel}

\usepackage{amstext}
\usepackage{graphicx}
\usepackage{amssymb}
\usepackage{esint}
\usepackage[unicode=true,pdfusetitle,
 bookmarks=true,bookmarksnumbered=false,bookmarksopen=false,
 breaklinks=false,pdfborder={0 0 1},backref=false,colorlinks=false]
 {hyperref}
\hypersetup{
 pdfauthor={Florian Bauer, Joan Sol\`a, Hrvoje \v{S}tefan\v{c}i\'{c}},
 pdfkeywords={Cosmological constant, modified gravity}
 }

\renewcommand{\[}{\begin{equation}}
\renewcommand{\]}{\end{equation}}

\newcommand{\rL}{\rho_{\Lambda}}
\newcommand{\rLi}{\rho_{\Lambda}^{i}}
\newcommand{\rLe}{\rho_{\Lambda{\rm eff}}}

\newcommand{\rLo}{\rho_{\Lambda}^0}
\newcommand{\CC}{\Lambda}

\newcommand{\newtext}[1]{\text{#1}}

\begin{document}

\title{Relaxing a large cosmological constant\\ in the astrophysical domain}

\author{{\bf Florian Bauer}\textsuperscript{a}, {\bf Joan Sol\`a}\textsuperscript{a},  {\bf Hrvoje
 \v{S}tefan\v{c}i\'{c}}\textsuperscript{b}\\
\\
\textsuperscript{a} HEP Group, Dept.\ ECM and Institut de Ci{\`e}ncies del Cosmos\\
Univ.\ de Barcelona, Av.\ Diagonal 647, E-08028 Barcelona,
Catalonia, Spain\\
\\
\textsuperscript{b} Theoretical Physics Division, Rudjer Bo\v{s}kovi\'{c}
Institute \\PO Box 180, HR-10002 Zagreb, Croatia\\ \\ emails:
fbauer@ecm.ub.es, sola@ecm.ub.es, shrvoje@thphys.irb.hr}

\date{}

\maketitle

\begin{abstract}
We study the problem of relaxing a large cosmological constant in
the astrophysical domain through a dynamical mechanism based on a
modified action of gravity previously considered by us at the
cosmological level. We solve the model in the
Schwarzschild-de~Sitter metric for large and small astrophysical
scales, and address its physical interpretation by separately
studying the Jordan's frame and Einstein's frame formulations of it.
In particular, we
determine the extremely weak strength of fifth forces
in our model and show that they are virtually unobservable. %
Finally, we estimate the influence that the relaxation
mechanism may have on pulling apart the values of the two
gravitational potentials $\Psi(r)$ and $\Phi(r)$ of the metric, as
this implies a departure of the model from General Relativity and
could eventually provide an observational test of the new framework
at large astrophysical scales, e.g.\ through gravitational lensing.

\end{abstract}



\section{Introduction\label{sec:Intro}}

The old cosmological constant (CC) problem\,\cite{weinberg89} is
considered as one of the biggest puzzles in theoretical physics.
Remarkably, it is only partially related to the question of what
drives the current accelerated expansion\,\cite{Observations1,Observations2,Observations3}. In
general the latter question is ``answered'' by the so-called ``dark
energy'' (DE), which summarizes a large amount of different models
and concepts\,\cite{CCproblem1,CCproblem2,CCproblem3,CCproblem4,CCproblem5,CCproblem6}. The simplest candidate for the DE is
a tiny positive cosmological constant $\CC$, others are quintessence
scalar fields, modified gravity, the existence of large scale
ambiguities or even a misinterpretation of observations, just to
mention a few of them. 
Whatever the true answer
might be, the old CC problem still remains in its full glory because
we cannot just put the huge vacuum energy predicted by every quantum
field theory (QFT) under the rug, and we cannot just resolve the
problem by mere fine tuning methods -- cf.\ e.g.\ the recent
references \cite{BSS-2009-01,BSS-2009-02,BSS-2009-03} and \cite{BSS-2010}\footnote{See
specially the detailed account of the fine tuning CC problem in
Appendix B of \cite{BSS-2010}.}. In essence, the CC problem stems
from the profound incompatibility between the observed DE
density\,\cite{Observations1,Observations2,Observations3} with the theoretical
expectations\,\cite{weinberg89,CCproblem1,CCproblem2,CCproblem3,CCproblem4,CCproblem5,CCproblem6}. One has to admit that
theoretical physics is currently not able to predict the value of
the vacuum energy density $\rL=\CC/(8\pi G)$. In actual fact, this
is not the main preoccupation, since after all QFT is unable to
predict, say, the value of the electron mass. The real problem
instead is quite another one, to wit: while renormalizable QFT does
provide a fully consistent framework to accommodate the value of the
electron mass (namely, one which is free from UV and IR
ambiguities), it is nevertheless unable to do the same (as far as we
know) for the measured value of $\rL$. As a result, we are left with
various rough order of magnitude estimates of finite vacuum energy
contributions, e.g.\ from phase transitions and inflation, in
addition to naive and mostly unreliable calculations of quantum
zero-point energies -- see however\,\cite{CC-BEC1,CC-BEC2,CC-BEC3} and references
therein. In this untidy state of affairs, there is at least one
single thing of which we can say that we are fully convinced to be
true: that the huge contributions to the CC come from theories that
work at typical energies much larger than those observed in the
present universe, and therefore any practical mechanism is bound to
start working effectively and efficiently already at these large
scales, and then persist in the job for a sufficiently large period
of time until our days.

A promising dynamical route along these lines, i.e.\ one which is
free from fine tuning problems, is taken by models which intend to
make a low-energy universe as ours feasible despite having a large
CC in the total energy content. In Refs.~\cite{BSS-2009-01,BSS-2009-02,BSS-2009-03,BSS-2010}
we proposed some intriguing models of this kind, where the
machinery at work was called the CC relaxation mechanism. The
resulting cosmos was referred to as the \emph{relaxed universe}. It
is not surprising that these more innovative approaches require more
invasive changes in the theory than just providing a source of
late-time acceleration, as it occurs in the usual modified gravity
models\,\cite{ModifiedGrav1-01,ModifiedGrav1-02,ModifiedGrav1-03,ModifiedGrav1-04,ModifiedGrav2-01,ModifiedGrav2-02,ModifiedGrav2-03,ModifiedGrav2-04}. While some of the latter
are well motivated by observations, they usually presume the
existence of an implicit (and extremely fine tuned) counterterm in
the effective action just devised to cancel the big CC, such that
the other sources (matter and the new terms of the modified
gravitational action) can drive the acceleration on top of the
essentially zeroed (by hand) vacuum energy. Hence, even the most
viable modified gravity models in the literature are in urgent need
of some kind of CC relaxation that can avoid the unacceptable fine
tuning of the vacuum energy which is implicit in all of them. The
enormous difficulty in providing some form of solution capable to
alleviate this acute problem plaguing all the traditional (i.e.\
``late-time'') models of modified gravity gives a strong motivation
to invoke new and unconventional concepts, with the understanding
that in the beginning they might have the status of toy models or
prototype ideas just hinting at the final solution. A serious hint
along these lines might just be to replace the ``late-time''
modification of gravity by an ``all time'' (i.e.\ perennial or
everlasting) modification of it.

In Ref.~\cite{BSS-2010} we discussed a specific perennial modified
form of gravity which provides an implementation of the CC
relaxation mechanism at the cosmological level using the FLRW
metric. We found that the evolution of the universe is well behaved
despite the presence of a huge initial vacuum energy density of
order $\rLi\sim M_X^4\lesssim M_P^4$ ($M_P$ being the Planck mass),
in fact much larger in magnitude than the energy density of matter
and radiation, thanks to the dynamical compensation of $\rLi$
triggered by the expansion itself \,\footnote{A detailed
confrontation of these kind of models with the modern cosmological
data has been recently presented in Ref.\,\cite{BBS-2011}. It is remarkable that they are able to fit the expansion history of the
universe in a comparable way to the concordance $\CC$CDM model.}. The fact
that this relaxation mechanism must be active during most periods of
cosmological evolution (actually, at \textit{all times} after the
inflationary period) shows the intriguing nature of this approach.
But it also explains the difficulty of generalizing the mechanism to
more complicated systems, specially those that unavoidably involve
the interplay of gravity and the matter sources. In order to
understand better the properties of the CC relaxation mechanism
developed in~\cite{BSS-2010}, we investigate here its behavior on
sub-horizon scales, actually on length scales fully in the
astrophysical domain, e.g.\ in the Solar system environment. Apart
from searching for adequate metric solutions in spherically
symmetric backgrounds, and dwelling on the physical interpretation
of our model in different gravity frames, we will estimate the
strength of the possible fifth forces. Indeed, in many modified
gravity theories these forces appear as a manifestation of extra
gravitational degrees of freedom~(\textit{d.o.f.}). The purpose of
this Letter is to obtain a qualitative understanding of all these
effects. Hence, we restrict the discussion to the special case of
static backgrounds. At the cosmological level this corresponds to
the de~Sitter solution we found in~\cite{BSS-2010}, which defines
the border line between the quintessence-like and phantom-like
behavior exhibited by our model in the future regime. Consistently,
we expect that the static solutions we deal with do not represent
completely stable configurations. However, since we have found
in~\cite{BSS-2010} that the cosmological future can be very close to
the de~Sitter cosmos at the present time, we may nevertheless obtain
useful information from the study of the static case without facing
the full time-dependent situation. Finally, we discuss some
numerical solutions leading to possible deviations with respect to
General Relativity.

\section{The relaxation mechanism at astrophysical scales\label{sec:Spherical-symmetric}}

The action describing our scenario is given by\,\cite{BSS-2010}
\begin{equation}
S=\int
d^{4}x\sqrt{|g|}\left[\frac{1}{2}M_{P}^{2}R-\left(\frac{\beta}{B}+
\frac{\beta_{\odot}}{R}\right)-\rho_{\Lambda}^{i}\right]+S_{\text{mat}}[g_{ab}]\label{eq:Action-Complete}
\end{equation}
with the standard Einstein-Hilbert term $(1/2)M_{P}^{2}R=R/(16\pi\,G_N)$,
the matter action~$S_{\text{mat}}$ and the large vacuum energy
term~$\rho_{\Lambda}^{i}$. The modifications to gravity are
described by terms proportional to $\beta$ and $\beta_{\odot}$, and
the function~$B=B(R,G)$ is given in terms of the Ricci scalar~$R$
and the Gau\ss-Bonnet
invariant~$G=R^2-4R_{ab}R^{ab}+R_{abcd}R^{abcd}$.

The corresponding gravitational field equations follow from the
functional variation of the action (\ref{eq:Action-Complete}). After
a straightforward calculation one finds:

\begin{equation}\label{eq:fieldequations}
R_{ab}-\frac12\,g_{ab}\,R= -8\pi G_{N}\left[T_{ab}+\rLi\,g_{ab}+2
E_{ab}\right]\,,
\end{equation}
where $T_{ab}$ is the ordinary energy-momentum tensor, and
$E_{ab}$ is the extra tensor that appears as a consequence of the
modified gravity terms. Explicitly,
\begin{eqnarray}
E_{ab} & = &\frac12
g_{ab}\left(\frac{\beta}{B}+\frac{\beta_{\odot}}{R}\right)
+\left(R_{ab}+\nabla_a\nabla_b-g_{ab}\Box\right)\left(\frac73\frac{\beta}{B^2}R+\frac{\beta_{\odot}}{R^2}\right)\label{eq:Eabfull}\nonumber\\
& + & 2\beta\left\{\frac{R_{amnl}R_b^{\,\,mnl}-4\,R_{ac}R_b^c}{2\,B^2}
+\left(g_{ab}\nabla_c\nabla_d-2\,g_{bd}\,
  \nabla_a\nabla_b\right)\left(\frac{R^{cd}}{B^2}\right)\right.\nonumber\\
& + & \left.\Box\left(\frac{R_{ab}}{B^2}\right)
-\nabla_m\nabla_n\left(\frac{R^n_{\,\,ab}{}^m}{B^2}\right)\right\}.
\end{eqnarray}
We see that this tensor follows solely from the gravity
modifications in (\ref{eq:Action-Complete}). If we set
$\beta=\beta_{\odot}=0$ in it, then ${E}_{ab}=0$ and
(\ref{eq:fieldequations}) boils down to the standard Einstein's
equations with non-vanishing vacuum energy density $\rLi$, i.e.\
$R_{ab}-(1/2)g_{ab}R=-8\pi G\,\left(T_{ab}+\rLi\,g_{ab}\right)$. In
the general case, the field equations are rather cumbersome and we
will have to consider different simplified situations.

A working ansatz for $B(R,G)$ is given in Ref.\cite{BSS-2010}, where
it is then specialized to the FLRW background and hence valid for
cosmological considerations. In this domain, and for all epochs
after the radiation epoch, it takes the simplified form
\begin{equation}\label{BRG}
B(R,G)=\frac{2}{3}R^{2}+\frac{1}{2}G\,,
\end{equation}
which is the precise combination that reproduces the correct
radiation and matter epochs through the cosmological
evolution\,\cite{BSS-2010}. Therefore, this expression also applies
for astrophysical applications, although in this case it must be
evaluated for a metric amenable to the new context, typically a
spherically symmetric one (see below).

On theoretical grounds, the initial CC density in the early universe
should read $\rLi\sim M_X^4$ (where $M_X$ can be some GUT scale near
$M_P$), and is to be taken much larger in magnitude than any other
energy density. Therefore, the~$M_{P}^{2}R$ and matter terms
$\rho_{\text{mat}}$ in the action (\ref{eq:Action-Complete}) are
much smaller than $\rLi$. Even if we take $\rLi$ as the vacuum
energy density of the Standard Model (SM) of Particle Physics, we
have $\rho_{\Lambda}^{i}\sim-(2\times 10^{2}\,\text{GeV})^{4}$,
which induces a very large cosmological constant, some 56 orders of
magnitude away from the measured value $\rLo\sim 10^{-47}$ GeV$^4$.
That (negative) contribution from the electro-weak phase transition
is probably the most reliable number we can assume at the moment for
the theoretical value of the vacuum energy, inasmuch as it
represents the prediction of the SM as the most successful QFT to
date. According to our earlier work \cite{BSS-2009-01,BSS-2009-02,BSS-2009-03,BSS-2010}, the
gravity modifications in (\ref{eq:Action-Complete}) can be used to
relax dynamically the large CC and leaving a low-curvature
cosmological solution which is not dominated
by~$\rho_{\Lambda}^{i}$. This is possible because the effective
vacuum energy density in our $F(R,G)$-cosmology -- and hence the
quantity playing the role of DE in our framework -- is \textit{not
just} the parameter $\rLi$, but the full expression
$\rLe=\rLi+2\,E^{0}_{0}$, as one can see from
(\ref{eq:fieldequations}). The effective quantity $\rLe$ is
dynamically enforced to be very small by the relaxation
mechanism\,\cite{BSS-2010}. If we focus on the relevant terms of the
action which are acted upon by this mechanism, the corresponding
field equations read
\begin{equation} 2E_{ab}+\rho_{\Lambda}^{i}\,
g_{ab}=0\,.\label{eq:SSS-EinsteinEqs}
\end{equation}
Solving this equation means that there is a dynamical cancelation
of~$\rho_{\Lambda}^{i}$ by the $\beta,\beta_{\odot}$-terms in
(\ref{eq:Eabfull}). The remaining terms play no role for the
relaxation because
\begin{equation}
\left|\frac{\beta}{B}+\frac{\beta_{\odot}}{R}\right|\sim\left|\rho_{\Lambda}^{i}\right|
 \gg M_{P}^{2}R,\,\rho_{\text{mat}}.\label{eq:betaTerms-large}
\end{equation}
The details of the relaxation mechanism and various numerical
examples in the cosmological context have been discussed at length
in the comprehensive paper~\cite{BSS-2010}.


For typical values of the $\beta$ parameters, the $\beta_{\odot}/R$
term has little influence on the large scale cosmological expansion
(as we shall discuss below). Therefore, the standard problems of
$1/R$ gravity (as e.g.\ the incorrect description of the matter and
radiation dominated epochs in these models\,\cite{Polarski07-01,Polarski07-02}) do
not apply in our case because at very large (cosmological) scales
the leading term in the action (\ref{eq:Action-Complete}) is not
$1/R$ but $1/B$, and we know that the latter works perfectly well
since it is capable of relaxing the large CC and it correctly
reproduces the standard radiation and matter dominated
epochs~\cite{BSS-2010}.

On small length scales, instead, it is the $1/R$ term of
(\ref{eq:Action-Complete}) that dominates, and therefore it will be
this term that is responsible for the CC relaxation in the
astrophysical domain. To better understand this behavior, let us
study static spherically symmetric solutions described by the
metric\begin{equation}
ds^{2}=a(r)dt^{2}-\frac{dr^{2}}{b(r)}-r^{2}\left(d\theta^{2}+\sin^{2}\theta\,
d\phi^{2}\right)\,,\label{eq:SSS-metric}\end{equation} with two
functions $a(r)$, $b(r)$ depending only on the radial
coordinate~$r$. The time and angle coordinates play no role in the
following. As mentioned already in Sec.~\ref{sec:Intro}, static
solutions on small scales correspond to the asymptotic de~Sitter
metric at the cosmological level separating the quintessence from
the phantom future behavior. Consequently, the static setup
considered here is not expected to be absolutely stable. However, it
may be a sufficiently useful approximation for obtaining some
insights into situations not far to the de~Sitter future. According
to the recent observations, which prefer a CC-like equation of state
for the DE, $\omega_D\simeq -1$, this a reasonable assumption.

In the general case, finding solutions of the relaxation equation
(\ref{eq:SSS-EinsteinEqs}) for the astrophysical problems can get
quite complicated, so further approximations are necessary. The
simplest situation that we wish to consider is the de~Sitter,
spherically symmetric, space-time. It corresponds to the form given
in Eq.\,(\ref{eq:SSS-metric}) with
\begin{equation}\label{eq:deSitter}
a(r)=1-\frac{r^{2}}{r_{e}^{2}},\,\,\, \ \ \ b(r)=a(r)\,,
\end{equation} thus representing a 3-dimensional sphere of radius
$r_{e}=\sqrt{3/\CC}$, which acts as a cosmological horizon. Such a
space-time  is an exact solution of (\ref{eq:SSS-EinsteinEqs}) for
all~$r$, as we can check. Let us indeed compute the extra tensor
$E_{ab}$ for the de Sitter metric. First of all we find
$R=12r_{e}^{-2}$ and $B=108r_{e}^{-4}$, where we used (\ref{BRG}) to
compute the latter. Finally, from (\ref{eq:Eabfull}) we arrive at:
\begin{equation}
E_{ab}=\left\{\beta\frac{r_{e}^{4}}{108}+\beta_{\odot}\frac{r_{e}^{2}}{16}\right\}\,g_{ab}\,.\label{eq:sss-E00-dS}\end{equation}
The fact that $E_{ab}$ is constant confirms that the de Sitter
space-time is an exact solution. Equation (\ref{eq:SSS-EinsteinEqs})
can then be satisfied and so the expression (\ref{eq:sss-E00-dS})
cancels against $\frac{1}{2}\rho_{\Lambda}^{i}\,g_{ab}$.
{As both terms with $\beta$ and $\beta_{\odot}$  are in general needed to cancel $\rho_{\Lambda}^{i}$, they together need to produce the contribution of the sign opposite to $\rho_{\Lambda}^{i}$. The straightforward way to achieve this for $\rho_{\Lambda}^{i}>0$ is to have both $\beta<0$ and $\beta_{\odot}<0$, whereas for $\rho_{\Lambda}^{i}<0$ to have both $\beta>0$ and $\beta_{\odot}>0$.}
Finally, after imposing the boundary condition on the current value of $\CC$, this
fixes the order of magnitude of these parameters\,\footnote{It
should be clear that the values of $\beta$ and $\beta_{\odot}$ need
not be fine-tuned at all, we only have to fix their order of
magnitude and sign -- see equations (\ref{reCosmological}) and
(\ref{reAstrophysical}) below. The relaxation mechanism then selects
$r_e$ automatically such that (\ref{eq:SSS-EinsteinEqs}) is
fulfilled, irrespective of the input value of
$\rLi$\,\cite{BSS-2010}}. Of course the cancelation is not exact,
leaving a very small remainder (the current value of $\rLo$), but
the mechanism does not depend on it. Let us emphasize that this
cancelation involves no fine tuning because $r_e$ is not a constant
to be fixed by us, but a dynamical variable controlled by the
relaxation mechanism. Whatever it be the starting value for $\rLi$,
the relaxation mechanism chooses $r_e$ such that
(\ref{eq:SSS-EinsteinEqs}) is fulfilled. Recall that
$r_{e}=\sqrt{3/\CC}=H^{-1}$ (with $\CC=8\pi G_N\,\rLe$) is driven to
the current value of order $H_{0}^{-1}$. The reason for this
dynamical choice stems from the late time epoch of the universe
evolution, in which the relaxation condition (viz.{} $B\to 0$, without
ever being exactly zero) enforces a very small value of
$H$\,\cite{BSS-2010}. Since $r_e$ becomes then very large, we see
from Eq.(\ref{eq:sss-E00-dS}) that the expression $E_{00}$ becomes
also very large, in fact as large as to essentially cancel against
$\rLi\sim M_X^4$. Notice that the first term of
(\ref{eq:sss-E00-dS}) behaves as $\sim r_e^4$ and hence dominates at
very large distances; this is the term that emerges from the
$1/B$-invariant and which we used in the cosmological domain to
insure a very small value for the measured CC\,\cite{BSS-2010}. As
we can see from (\ref{eq:sss-E00-dS}), in this domain $r_e$ is
dynamically driven to satisfy $\beta\,r_e^4/54+\rLi=\rLo\ll\rLi$,
i.e.\ effectively Eq.\,(\ref{eq:SSS-EinsteinEqs}). Therefore,
\begin{equation}\label{reCosmological}
r_e\sim\left|\frac{\rLi}{\beta}\right|^{1/4}\sim
\frac{M_X}{{\cal M}^2}\,,
\end{equation}
where we have set $|\beta|\equiv {\cal M}^8$. Choosing ${\cal M}=\sqrt{M_X\,H_0}$
we get $r_e\sim H_0^{-1}$. For $M_X\sim 10^{16-17}\,\text{GeV}$ (GUT/string
scale), we see that the required order of magnitude for ${\cal M}$ is in
the $\,\sim$meV range, characteristic of a light neutrino mass, and
is also the energy scale of the CC: $m_{\CC}\equiv
\left(\rLo\right)^{1/4}\sim 10^{-3}$ eV.

The relaxation mechanism also works fairly well in the astrophysical
domain. It only changes qualitatively, as we shall discuss below,
where now it is the $1/R$ term of the action
(\ref{eq:Action-Complete}) that takes over. But the relaxation
mechanism once more works appropriately to protect the astrophysical
scales from unwanted vacuum effects. To start with, we note that the
astrophysical domain cannot be described by just the spherically
symmetric de Sitter metric (\ref{eq:deSitter}); we need the
Schwarzschild component too. Let us therefore consider the
Schwarzschild-de~Sitter (SdS) ansatz with non-zero Schwarzschild
radius~$r_{s}\ll r_{e}$:\begin{equation}
a(r)=b(r)=1-\frac{r^{2}}{r_{e}^{2}}-\frac{r_{s}}{r}.\label{eq:SdS-Func-a}\end{equation}
In contrast to standard General Relativity, this metric is not a
solution of our system because the Gau\ss-Bonnet term in $\beta/B$
induces an $r$-dependence in the extra tensor $E_{ab}$
in~(\ref{eq:SSS-EinsteinEqs}), which cannot be compensated by the
constant value of $\rLi$. However, for sufficiently small or large
values of the radius, $E_{ab}$ is still approximately constant. To
better understand this issue, let us look at the extra tensor
following from this metric. In the astrophysical setup, the
invariant functions in the action~(\ref{eq:Action-Complete}) take on
the form
\begin{equation}\label{denomRB}
~R=12r_{e}^{-2}\,,\ \ \ \ \ \ \ \ \ B=108r_{e}^{-4}+6r_{s}^{2}\cdot
r^{-6}\,.
\end{equation}
Notice that the second term of $B$ in (\ref{denomRB}) is the only
new effect that the Schwarzschild geometry introduces on the
previous result for the de Sitter case. The correction just comes
from the Gau\ss-Bonnet invariant part of $B$ in (\ref{BRG}), which
reduces to the square Riemann tensor
$G=R_{abcd}R^{abcd}=12\,r_s^2/r^6$ for the Schwarzschild geometry,
and induces the aforesaid $r$-dependence. From the expression of $B$
in (\ref{denomRB}), we see that in the limit of very large radius,
$r\rightarrow r_{e}$,  the leading term of (\ref{eq:Action-Complete})
that implements the relaxation mechanism is still the $\beta/B$ one,
and this fixes $\beta$ to be of order $\beta\sim
\rLi/r_e^4\sim\rLi\,H_0^4$, as we have discussed in
(\ref{reCosmological}). Thus, in this large scale regime the
${\beta_{\odot}}/{R}$ term plays no significant role. However, for
local astrophysical scales the situation changes and then it is the
$1/R$ term which takes over. Unfortunately, a detailed discussion of
the SdS geometry is difficult. Even in the case where
$\beta_{\odot}$ is set to zero, the field equations
(\ref{eq:SSS-EinsteinEqs}) for the relaxation mechanism become quite
complicated owing to the $r$-dependence of $B$. Albeit this case
will be treated numerically in section
\ref{sec:Numerical-results-betaB}, some qualitative considerations
will be helpful to better grasp the behavior of the CC relaxation
mechanism in the local astrophysical domain. In this simple context
we can keep both the $\beta$ and $\beta_{\odot}$ terms.

From the structure of the extra tensor $E_{ab}$ in
(\ref{eq:Eabfull}), one can see that the $R$ and $B$ invariants
appear in the denominator of~$E_{ab}$ with some powers,
schematically\[
E_{00}=\beta_{\odot}\frac{f_{1}(r)}{R^{n_{1}}}+\beta\frac{f_{2}(r)}{B^{n_{2}}}\,,\]
with $n_{1},n_{2}>0$. Ignoring the details of $f_{1,2}$ and
$n_{1,2}$ we may use the simple form \[
E_{00}=\frac{3}{4}\frac{\beta_{\odot}}{R}+\frac{\beta}{B}=\frac{\beta_{\odot}}{16r_{e}^{-2}}+
\frac{\beta}{108r_{e}^{-4}+6r_{s}^{2}\cdot r^{-6}}\,,\] which should
suffice to obtain some qualitative results. For $r\rightarrow\infty$
we recover the de~Sitter result in Eq.~(\ref{eq:sss-E00-dS}), which
is an exact solution. For smaller values of~$r$, the non-constant
term $6r_{s}^{2}\cdot r^{-6}$ starts to become more important and it
will eventually dominate in~$\beta/B$, specifically
for~$r<r_{c1}:=(r_{s}^{2}r_{e}^{4}/18)^{1/6}$. In this region the
SdS metric is not a good approximate solution of
(\ref{eq:SSS-EinsteinEqs}).

Finally, for small~$r$ in the local astrophysical domain, we have
$1/B\propto r^6$ and the influence of this term wanes in comparison to
the $r$-independent contribution $1/R\sim r_e^2$, which then takes
over. Indeed, for small enough $r$, we find\begin{equation}
E_{ab}\approx
\left(\beta_{\odot}\frac{r_{e}^{2}}{16}+\beta\frac{r^{6}}{6r_{s}^{2}}\right)\,g_{ab}\to
\beta_{\odot} \frac{r_{e}^{2}}{16}\,g_{ab}\ \ \ \ {\rm for}\ \ \ \
r\ll
r_{c2}:=\left(r_{s}^{2}r_{e}^{2}\frac{6\beta_{\odot}}{16\beta}\right)^{1/6}\,.\label{eq:SdS-r-small}\end{equation}
In the local domain $ r\ll r_{c2}$, the tensor $E_{ab}$ is
approximately constant and the equations (\ref{eq:SSS-EinsteinEqs})
are again satisfied for an appropriate dynamical choice of $r_e$.
Thus, the SdS metric becomes an acceptable solution for dynamical
relaxation in this region, too. From (\ref{eq:SdS-r-small}) it
follows that the CC can be successfully relaxed in the astrophysical
domain provided  $r_e$ is dynamically driven to fulfill the relation
$\beta_{\odot}\,r_e^2/8+\rLi=\rLo$, which again in practice means
$\rLi+\beta_{\odot}\,r_e^2/8=0$, i.e.\
Eq.\,(\ref{eq:SSS-EinsteinEqs}). In any case, we conclude
\begin{equation}\label{reAstrophysical}
r_e\sim\left(\frac{\rLi}{|\beta_{\odot}|}\right)^{1/2}\sim
\frac{M_X^2}{{\cal M}'^3}\,,
\end{equation}
where $|\beta_{\odot}|\equiv {\cal M}'^6$. Compare (\ref{reAstrophysical})
with the corresponding result in the cosmological domain,
(\ref{reCosmological}). Choosing ${\cal M}'^3=M_X^2\,H_0$ we get once more
$r_e\sim H_0^{-1}$. In this case, however, for $M_X\sim
10^{16-17}\,\text{GeV}$, the new (astrophysical) scale ${\cal M}'$ must be located in the MeV range, which is a common scale for the SM of Particle
Physics. It suggests a possible natural connection of the required
order of magnitude value of the parameter
$\beta_{\odot}=-8\rLi\,r_e^{-2}\simeq -\rLi\,H_0^2$ with
conventional physics. Therefore, we could venture a possible natural
explanation of why $r_e$ is so large, also in the astrophysical
domain: it might be only because $M_X$ is much larger than the
typical particle physics scales in the SM. If so, the hierarchy
problem in Particle Physics could be linked to the hierarchy of
vacuum energy scales in astrophysics and cosmology.  \newtext{However, this requires further studies.} Far beyond the local astrophysical scales, we have the intermediate domain corresponding to the intergalactic and the intercluster distances, which is more complicated to analyze. See Sec.~\ref{sec:Numerical-results-betaB} for a preliminary study of that cosmological intermediate region.

\section{Studying the model in different gravity frames\label{sec:Fifth-forces}}

In this section, we investigate the existence and possible impact of
extra \textit{d.o.f.}\ in the relaxation model at small length
scales, e.g.\ in the Solar system environment. In many theories of
modified gravity new \textit{d.o.f.}\ show up, which could mediate
an extra force on matter coupled to gravity. A common way to
identify the new \textit{d.o.f.}\ from the modified gravity action
is by considering its equivalent conformal scalar field
representation. Using the arguments from
Sec.~\ref{sec:Spherical-symmetric}, gravity in this regime can be
described in good approximation by a modified $F(R)$
theory\,\cite{ModifiedGrav2-01,ModifiedGrav2-02,ModifiedGrav2-03,ModifiedGrav2-04}. Indeed, the term~$\beta/B$ involving
the Gau\ss-Bonnet scalar~$G$ is neglected and the the extra
\textit{d.o.f.}\ is just one scalar field. Furthermore, since our
relaxation model is not just General Relativity with small
corrections, it is convenient to discuss the situation in more
detail. In particular, we will cross-check our results by
considering a direct expansion of the $F(R)$ theory around a
constant background and then identifying the new \textit{d.o.f.}

\subsection{Jordan frame}

The action~(\ref{eq:Action-Complete}) is formulated in the Jordan
frame, where matter is minimally coupled to gravity, i.e.\ to the
metric $g_{ab}$. With~$\beta=0$ we write it as a pure~$F(R)$ theory
plus matter.\begin{eqnarray}
S & = & \frac{1}{2}M_{P}^{2}\int
d^{4}x\sqrt{|g|}\left(\frac{1}{2}M_{P}^{2}R-\frac{\beta_{\odot}}{R}-\rho_{\Lambda}^{i}\right)+S_{\text{mat}}[g_{ab}]\\
& \equiv &
\frac{1}{2}M_{P}^{2}\int
d^{4}x\sqrt{|g|}F(R)+S_{\text{mat}}[g_{ab}],\label{eq:Action-FofR}
\end{eqnarray}
where we have defined the reduced Planck mass $M_{P}=1/\sqrt{8\pi
G}\sim10^{18}\,\text{GeV}$. As in Sec.~\ref{sec:Spherical-symmetric}
we use the SdS metric~(\ref{eq:SdS-Func-a}) as a starting point,
which allows fixing the parameter~$\beta_{\odot}$. Then, the
Einstein equations, emerging from $2\delta S/\delta g^{ab}=0$, are
approximately given by\begin{equation}
\left[-3M_{P}^{2}r_{e}^{-2}+\frac{\beta_{\odot}}{8}r_{e}^{2}+\rho_{\Lambda}^{i}\right]g_{ab}+T_{ab}=0,\label{eq:Einstein-Eqs}\end{equation}
where $T_{ab}$ follows from $2\delta(S_{\text{mat}})/\delta g^{ab}$
and $R=12r_{e}^{-2}\approx12H_{0}^{2}$. The equations
(\ref{eq:Einstein-Eqs}) are approximate since we keep the matter
content and at the same time we use the de Sitter value of the
curvature, which is justified inasmuch as we may use the de Sitter
solution as an useful approximation to the current universe.
Indeed, if we would strictly apply the condition
(\ref{eq:betaTerms-large}), then the
equations\,(\ref{eq:Einstein-Eqs}) boil down to
(\ref{eq:SSS-EinsteinEqs}), with $E_{ab}$ given by
(\ref{eq:SdS-r-small}). We discuss the stability issue in
Sec.~\ref{sub:Stability}. Note that from Eq.~(\ref{eq:Einstein-Eqs})
it follows that small changes in~${\beta}_{\odot}$ (or
$\rho_{\Lambda}^{i}$) yield only small changes in~$r_{e}$ and the
solution does not change much. Thus, while in our case $\rLi$ is
canceled by a dynamical choice of $r_e$ in which the matter sources
do not participate in a significant way, in standard General
Relativity the large value of $\rho_{\Lambda}^{i}$ must be canceled
\textit{ab initio} by a very precisely chosen counterterm, otherwise
the solution would change drastically, spoiling completely the
observed cosmological evolution. This observation displays the whole
dilemma of the old CC problem, and points towards a further
completion and improvement of the relaxation model.

Next, following the standard procedure for converting an $F(R)$ modified
gravity model into a scalar-tensor theory, we introduce an
auxiliary scalar field~$\phi$ by writing the
action~(\ref{eq:Action-FofR}) as\[ S=\frac{1}{2}M_{P}^{2}\int
d^{4}x\sqrt{|g|}\left[F(\phi)+F^{\prime}(\phi)(R-\phi)\right]+S_{\text{mat}}[g_{ab}].\]
The original action can be recovered from the variational principle
$\delta S/\delta\phi=0$, which yields~$\phi=R$ if
$F^{\prime\prime}(\phi)\neq0$. For the model given in
(\ref{eq:Action-FofR}), we have \[
F(\phi)=\phi-\frac{\tilde{\beta}_{\odot}}{\phi}-\frac{2\rho_{\Lambda}^{i}}{M_{P}^{2}},\,\,\,\
\ F^{\prime}(\phi)=1+\frac{\tilde{\beta}_{\odot}}{\phi^{2}},\] where
$\tilde{\beta}_{\odot}=2\beta_{\odot}/M_{P}^{2}$. In the SdS
solution, where $R=\phi=\phi_{s}\equiv 12r_{e}^{-2}$, the
dimensionless value of $F^{\prime}(\phi_{s})$ reads
\begin{equation}
F^{\prime}(\phi_{s})\approx
\frac{2(-8r_{e}^{-2}\rho_{\Lambda}^{i})}{(12r_{e}^{-2})^{2}M_{P}^{2}}=
-\frac{\rho_{\Lambda}^{i}}{3\rho_{c}^{0}},\label{eq:FprimeSdS}
\end{equation}
with the critical energy density
$\rho_{c}^{0}=3H_0^2/(8\pi\,G_N)=3r_{e}^{-2}M_{P}^{2}$ at present.
Obviously, $|F^{\prime}(\phi_{s})|\gg1$ is much larger in the
relaxation scenario than in more common~$F(R)$
models\,\cite{ModifiedGrav2-01,ModifiedGrav2-02,ModifiedGrav2-03,ModifiedGrav2-04}, where General Relativity is slightly
amended by a small correction.

\subsection{Einstein frame}

To obtain a standard Einstein-Hilbert term plus a scalar field
responsible for the fifth force we apply the conformal
transformation\begin{equation}
F^{\prime}(\phi)g_{ab}=\tilde{g}_{ab},\label{eq:ConfTrafo-b}\end{equation}
which takes us from the Jordan to the Einstein frame.

Within the Einstein frame, endowed with the metric $\tilde{g}_{ab}$
and corresponding curvature scalar~$\tilde{R}$, the transformed
action reads \[ S=\frac{1}{2}M_{P}^{2}\int
d^{4}x\sqrt{|\tilde{g}|}\left[\tilde{R}+\frac{3}{2}\tilde{g}^{ab}\frac{\nabla_{a}F^{\prime}\nabla_{b}F^{\prime}}{(F^{\prime})^{2}}-\frac{1}{(F^{\prime})^{2}}\left(\phi
F^{\prime}-F(\phi)\right)\right]+S_{\text{mat}}\left[\frac{\tilde{g}_{ab}}{F^{\prime}}\right]\,.\]
The canonical kinetic term for the scalar~$\varphi$ results from the
field redefinition $\phi\to\varphi$ as follows:
\begin{equation}\label{eq:defvarphi}
F^{\prime}(\phi)=e^{x}:=\exp\left(\sqrt{\frac{2}{3}}\frac{\varphi}{M_{P}}\right)\,.\end{equation}
Therefore, the action with canonically normalized fields is given by
\begin{equation} S=\int
d^{4}x\sqrt{|\tilde{g}|}\left[\frac{1}{2}M_{P}^{2}\tilde{R}+\frac{1}{2}\tilde{g}^{ab}\,\partial_{a}\varphi\,\partial_{b}\varphi-\frac{M_{P}^{2}}{2(F^{\prime})^{2}}\left(\phi
F^{\prime}-F(\phi)\right)\right]+S_{\text{mat}}\left[\frac{\tilde{g}_{ab}}{F^{\prime}}\right],\label{eq:Action-EF}\end{equation}
from which the the scalar potential can be read off, and its value
for the previous SdS solution be derived:
\begin{equation}
V(\varphi)=\frac{M_{P}^{2}\left(\phi
F^{\prime}(\phi)-F(\phi)\right)}{2(F^{\prime})^{2}}=\frac{M_{P}^{2}}{(F^{\prime})^{2}}\left(\frac{\tilde{\beta}_{\odot}}{\phi}+\frac{\rho_{\Lambda}^{i}}{M_{P}^{2}}\right),
\ V(\varphi_s)
=\frac{3M_{P}^{2}r_{e}^{-2}}{F^{\prime}(\phi_s)}
=\frac{\rho_c^0}{F^{\prime}(\phi_s)}\,,\label{eq:Pot-varphi}
\end{equation}
with $\phi=\phi_{s}=12r_{e}^{-2}$  (associated to the value
$\varphi_{s}$ of $\varphi$ in (\ref{eq:defvarphi})),
$\rLi=-9\,r_e^{-2}\,M_P^2\,F^{\prime}(\phi_s)$, and moreover
$\tilde{\beta}_{\odot}=(12r_{e}^{-2})^{2}\,F^{\prime}(\phi_s)$.
Obviously, $V(\varphi_{s})$ is the vacuum energy density in Einstein
frame variables (coincident to the critical density in that frame,
in the de Sitter approximation).

\subsection{Stability\label{sub:Stability}}

For studying the stability of the SdS solution we have to analyze
the potential in~(\ref{eq:Pot-varphi}), rewritten as
\begin{equation}
V(\varphi)=M_{P}^{2}\sqrt{\tilde{\beta}_{\odot}}\left(\sqrt{e^{x}-1}+\frac{\rho_{\Lambda}^{i}}{\sqrt{\tilde{\beta}_{\odot}}M_{P}^{2}}\right)e^{-2x}\simeq
M_{P}^{2}\sqrt{\tilde{\beta}_{\odot}}\left(e^{\frac{x}{2}}-\frac{3}{4}e^{\frac{x_{s}}{2}}\right)e^{-2x}\,,
\label{eq:Vofx}
\end{equation}
where from~(\ref{eq:FprimeSdS}) we have $|e^{x}|\gg1$ around the SdS
solution. Moreover,
$\tilde{\beta}_{\odot}=(12r_{e}^{-2})^{2}e^{x_{s}}$, in which
$e^{x_{s}}=F^{\prime}(\phi_{s})$ according to (\ref{eq:defvarphi}). %
\begin{figure}
\centering{}\includegraphics[clip,width=0.9\columnwidth]{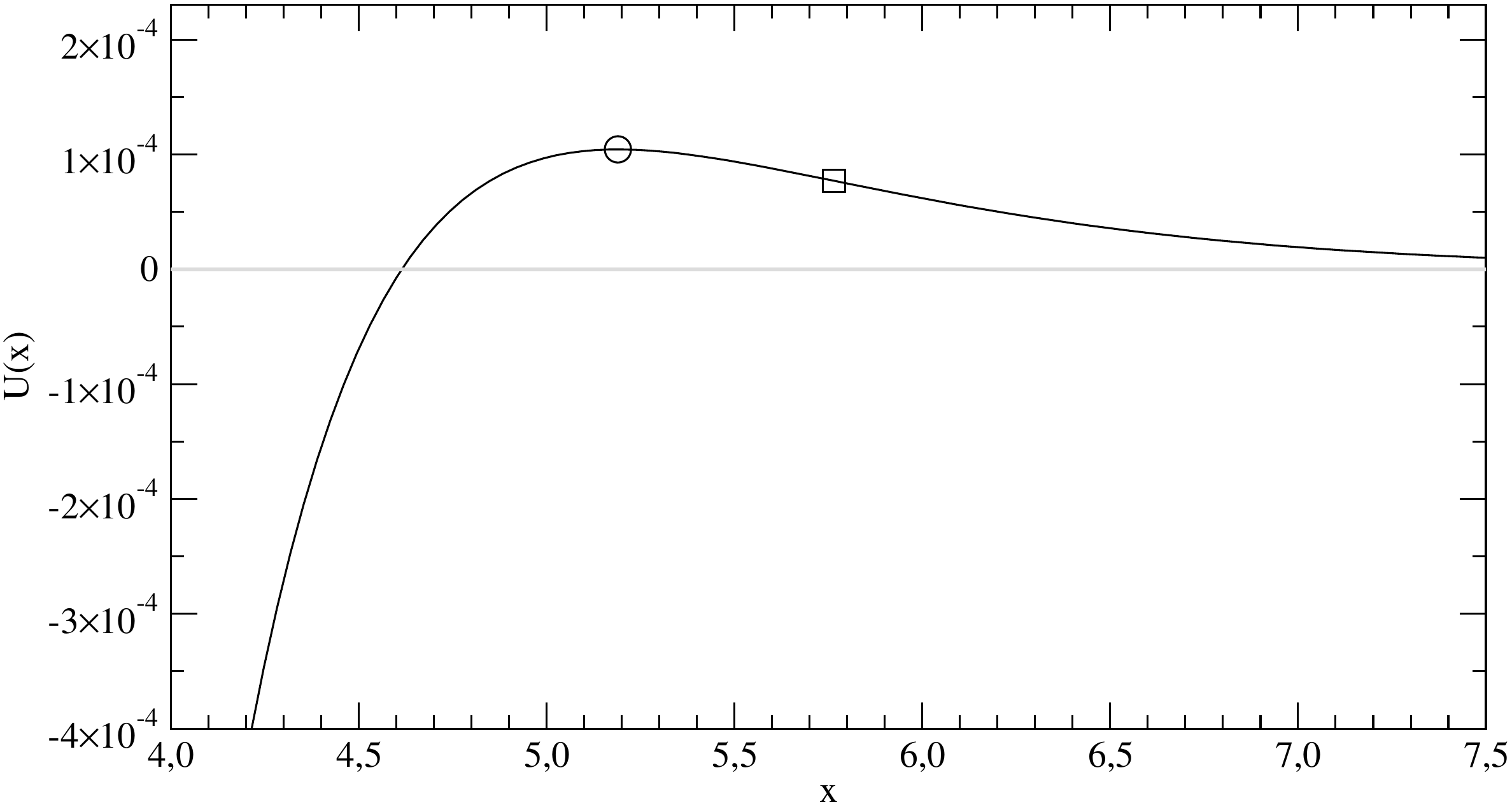}
\caption{Scalar field potential $U(x):=V(x)/(M_{P}^{2}\sqrt{\tilde{\beta}_{\odot}})$
from Eq.~(\ref{eq:Vofx}) with $x=\sqrt{\frac{2}{3}}\varphi/M_{P}$. Here, $x_s$ was chosen such that $U(0)=-\frac{3}{4}e^{x_{s}/2}=-10$. The circle is at the maximum
and the square at the inflection point of~$U(x)$. To the right of
the latter the scalar mass squared is positive. Note that both
limits $U(x\rightarrow0)$ and
$U(x\rightarrow\infty)\rightarrow0$ are valid only for the pure
$\beta_{\odot}/R$ model since the effects of the $\beta/B$ term will
be important there.\label{fig:VofVarphi}}
\end{figure}
Thus, within the same approximation, \begin{equation}
V^{\prime}(\varphi)=-\sqrt{\frac{3}{2}}M_{P}\sqrt{\tilde{\beta}_{\odot}}\left(e^{\frac{x}{2}}-e^{\frac{x_{s}}{2}}\right)e^{-2x},\
\ \
V^{\prime\prime}(\varphi)=2\sqrt{\tilde{\beta}_{\odot}}\left(\frac{3}{4}e^{\frac{x}{2}}-e^{\frac{x_{s}}{2}}\right)e^{-2x}.\label{eq:VppOfVarphi}\end{equation}
Exactly at the SdS solution $x=x_{s}$ we obtain
$V^{\prime}(\varphi_{s})=0$ and
$V^{\prime\prime}(\varphi_{s})=-6r_{e}^{-2}/e^{x_{s}}<0$, indicating
that the SdS solution is actually a maximum, see
Fig.~\ref{fig:VofVarphi}. Noting that a scalar mass term in the
Einstein frame transforms as
$\sqrt{|\tilde{g}|}\,m_{\varphi}^2\,\varphi^2\to
(F^{\prime})^{2}\sqrt{|{g}|}\
m_{\varphi}^2\,(\varphi^2/F^{\prime})=\sqrt{|{g}|}\
(F^{\prime}\,m_{\varphi}^2)\,\varphi^2$ into the Jordan frame, it
follows that the mass squared of $\varphi$ in the Jordan's frame is,
in our case, $m_{\varphi,J}^2=-6r_{e}^{-2}$. The fact that it is
negative is not a surprise, it was expected from the aforementioned
fact that at the cosmological level the de~Sitter solution is a
maximum and hence unstable\,\cite{BSS-2010}, its lifetime being of
order~$H_{0}^{-1}$. However, when $e^{x/2}$ moves to larger values
than $\frac{4}{3}e^{x_{s}/2}$, then $m_{\varphi,J}^2$ becomes
positive and acquires physical meaning. From then on the scalar
field $\varphi$ behaves like decaying quintessence. At the
inflection point the vacuum energy
$V(e^{x/2}=\frac{4}{3}e^{x_{s}/2})=({189}/{256})\,V(\varphi_{s})$ is
of the same order of magnitude as the de~Sitter cosmological
constant. Thus the relaxation mechanism works also in the stable
quintessence regime.

It is interesting to reconfirm the above Jordan's frame calculation
of the mass squared of the new gravitational \textit{d.o.f.} by
expanding around the constant background Ricci scalar
$R_{0}=12r_{e}^{-2}$. Following the general procedure
of\,\cite{Chiba:2006jp-01,Chiba:2006jp-02}, we obtain the vacuum Klein-Gordon equation
in the Jordan frame, namely
$\left(\Box+m_{R_1}^{2}\right)\,R_{1}=0$, where the gravitational
\textit{d.o.f.} is denoted now by $R_{1}(r)$, and $m_{R_1}^{2}$ its
mass squared: \[
m_{R_1}^{2}=\frac{\left[F^{\prime}(R_{0})-R_{0}F^{\prime\prime}(R_{0})-3\square
F^{\prime\prime}(R_{0})\right]}{3F^{\prime\prime}(R_{0})}=-\frac{1}{2}R_{0}=-6r_{e}^{-2},\]
and we used $F(R)=-{\tilde{\beta}_{\odot}}/{R}$,
$F^{\prime}(R)={\tilde{\beta}_{\odot}}/{R^{2}}$,
$F^{\prime\prime}(R)=-{2\tilde{\beta}_{\odot}}/{R^{3}}$. We note
that $m_{R_1}^{2}$ exactly coincides with $m_{\varphi,J}^2$, which
confirms our expectations. From this alternative approach, we most
clearly see that the large value
of~$\tilde{\beta}_{\odot}\propto\rho_{\Lambda}^{i}$ does not enter
the mass calculation, which means that the result does not depend
neither on the size nor on the sign of $\rho_{\Lambda}^{i}$. This is
different from the previously considered $\mu^{4}/R$ or $R^2/\mu^2$
models in the
literature~\cite{ModifiedGrav2-01,ModifiedGrav2-02,ModifiedGrav2-03,ModifiedGrav2-04,Polarski07-01,Polarski07-02,Chiba:2006jp-01,Chiba:2006jp-02}, where the
mass is directly related to the parameter~$\mu$.

\subsection{Coupling to matter\label{sub:Coupling-to-matter}}

Let us consider first the influence of matter on the evolution of
$\varphi$, and leave for the next section the study of the role of
$\varphi$ in the field equation of matter. The corresponding
Einstein's frame action reads\[ S=\int
d^{4}x\sqrt{|\tilde{g}|}\left[\frac{1}{2}M_{P}^{2}\tilde{R}+
\frac{1}{2}\tilde{g}^{ab}\,\partial_{a}\varphi\,\partial_{b}\varphi-V(\varphi)+\frac{1}{\sqrt{|\tilde{g}|}}\mathcal{L}_{\text{mat}}[g_{ab}]\right]\,,\]
where $g_{ab}=\tilde{g}_{ab}/(F^{\prime})$ is the
$\varphi$-dependent Jordan frame metric, and
$\mathcal{L}_{\text{mat}}$ is the matter Lagrangian density still in
the original Jordan's frame variables and containing the density
factor~$\sqrt{|g_{ab}|}$ in the Jordan metric. First, we look at the
Einstein equations following from $2\delta
S/\delta\tilde{g}_{ab}=0$,\begin{equation}
M_{P}^{2}\tilde{G}^{ab}+\tilde{T}_{\varphi}^{ab}+\frac{2}{\sqrt{|\tilde{g}|}}
\frac{\delta\mathcal{L}_{\text{mat}}[g_{ab}]}{\delta\tilde{g}_{ab}}=0,\label{eq:EF-Einstein}\end{equation}
where~$\tilde{T}_{\varphi}^{ab}$ is the standard energy-momentum
tensor of the canonically normalized scalar field including
~$V(\varphi)$. The matter energy-momentum tensor
reads\begin{equation}
\tilde{T}_{\text{mat}}^{ab}=\frac{2}{\sqrt{|\tilde{g}|}}\frac{\delta\mathcal{L}_{\text{mat}}[g_{ab}]}{\delta\tilde{g}_{ab}}=
\frac{2}{\sqrt{|\tilde{g}|}}\frac{\delta\mathcal{L}_{\text{mat}}[g_{ab}]}{\delta
g_{ab}}\frac{\partial
g_{ab}}{\partial\tilde{g}_{ab}}=\frac{2}{\sqrt{|\tilde{g}|}}\frac{\delta\mathcal{L}_{\text{mat}}[g_{ab}]}{\delta
g_{ab}}\frac{1}{F^{\prime}},\label{eq:Tmat}\end{equation} which
depends on~$\varphi$, too. Next we discuss the equations of motion
for~$\varphi$, given by
\begin{equation}
\frac{\delta
S}{\delta\varphi}=-\tilde{\square}\varphi-V^{\prime}(\varphi)+\frac{1}{\sqrt{|\tilde{g}|}}\frac{\delta\mathcal{L}_{\text{mat}}}{\delta
\varphi}=0,\label{eq:EF-varphi}\end{equation} where the variational
derivative in the last term can be determined with the help of
Eq.~(\ref{eq:Tmat}),\[
\frac{\delta\mathcal{L}_{\text{mat}}[g_{ab}]}{\delta\varphi}=\frac{\delta\mathcal{L}_{\text{mat}}[g_{ab}]}{\delta
g_{ab}}\frac{\partial
g_{ab}}{\partial\varphi}=\frac{\sqrt{|\tilde{g}|}}{2}F^{\prime}\tilde{T}_{\text{mat}}^{ab}\tilde{g}_{ab}\frac{\partial
F^{\prime}}{\partial\varphi}\frac{-1}{(F^{\prime})^{2}}=-\frac{\sqrt{|\tilde{g}|}}{\sqrt{6}\,M_P}\,\tilde{T}_{\text{mat}}^{ab}\tilde{g}_{ab}\,,\]
and in the last equality we used (\ref{eq:defvarphi}). Thus
Eq.\,(\ref{eq:EF-varphi}) can be cast as
$\tilde{\square}\varphi+V^{\prime}(\varphi)+\tilde{T}_{\text{mat}}^{ab}\tilde{g}_{ab}/\left(\sqrt{6}M_{P}\right)=0$,
where $\varphi$ is seen to couple to the trace of
$\tilde{T}_{\text{mat}}^{ab}$. The relation with the corresponding
energy-momentum tensor in the Jordan frame follows from
(\ref{eq:Tmat}):
$\tilde{T}_{\text{mat}}^{ab}=\sqrt{|{g}|/|\tilde{g}|}\,{T}_{\text{mat}}^{ab}/F^{\prime}=\left(F^{\prime}\right)^{-3}\,{T}_{\text{mat}}^{ab}$,
and hence the traces in both frames are related by $\
\tilde{g}_{ab}\tilde{T}_{\text{mat}}^{ab}=
\left(F^{\prime}\right)^{-2}\,g_{ab}\,{T}_{\text{mat}}^{ab}$.
Finally, since ${T}_{\text{mat}}^{ab}g_{ab}=\rho_{\text{mat}}$
provides the physical (non-relativistic) matter density in the
Jordan frame, we arrive at the effective field equation for
$\varphi$:
\begin{equation}
\tilde{\square}\varphi+V^{\prime}(\varphi)+\frac{1}{\sqrt{6}}\,\left(\frac{\rho_{\text{mat}}}{M_{P}\,{F^{\prime}}^{2}}\right)=0\,.\label{eq:VarphiEOM}\end{equation}
We may now compare the two source terms driving the evolution of
$\varphi$, viz.{} $V^{\prime}(\varphi)$ and
$\rho_{\text{mat}}/(M_P{F^{\prime}}^{2})$. If we look at the
beginning of the quintessence regime ($\varphi\gtrsim\varphi_{s}$),
where $\rho_{\text{mat}}\sim\rho_{c}^{0}=3M_{P}^{2}r_{e}^{-2}$, the
first source is of order
$V^{\prime}(\varphi)\sim(M_{P}\,r_{e}^{-2})/F^{\prime}\sim
\rho_{c}^{0}/(M_P\,F^{\prime})$, whereas the second source is of
order $\sim \rho_{c}^{0}/(M_P{F^{\prime}}^{2})$ and hence much
smaller owing to the additional suppression factor of $F^{\prime}$.
This shows that the evolution of~$\varphi$ is dominated by
$V(\varphi)$. Hence, matter does not disturb the relaxation
mechanism. This is not surprising because it is related to the
dominance of the curvature by the relaxation mechanism. Indeed, if
we take the trace of Eq.~(\ref{eq:EF-Einstein}), we obtain
$M_{P}^{2}\tilde{R}=\tilde{T}_{\varphi}^{ab}\tilde{g}_{ab}+\tilde{T}_{\text{mat}}^{ab}\tilde{g}_{ab}$.
From (\ref{eq:Pot-varphi}) we see that the first term on the
righthand side of the trace is of the order~$V(\varphi_{s})\sim
\rho_c^0\,(F^{\prime})^{-1}$ during the slow-roll regime, whereas
the second term is $\sim \rho_c^0\,(F^{\prime})^{-2}$, which is once
more smaller than $V(\varphi_{s})$ by a factor of
$(F^{\prime})^{-1}$.

\subsection{Possible fifth forces with matter}\label{sub:fifforce}

In the previous section we found that the evolution of the
gravitational scalar~$\varphi$ is not disturbed by matter.
Conversely, now we wish to study if matter can be significantly
affected by the presence of $\varphi$. We will illustrate it by
considering the effect on a matter scalar field~$\psi$ with
mass~$m_{\psi}$ in the Jordan frame. The corresponding matter action
reads
\[ S_{\text{mat}}[g_{ab},\psi]=\int
d^{4}x\sqrt{|g|}\left[\frac{1}{2}g^{ab}\,\partial_{a}\psi\,\partial_{b}\psi-\frac{1}{2}m_{\psi}^{2}\psi^{2}\right].\]
From the conformal transformation~(\ref{eq:ConfTrafo-b}), and the
associated relation $g^{ab}=(F^{\prime})\,\tilde{g}^{ab}$, we switch
into the Einstein frame. The complete action in this frame is given
in Eq.~(\ref{eq:Action-EF}), where in the present case the matter
part reads\[ \tilde{S}_{\text{mat}}\equiv
S_{\text{mat}}\left[\frac{\tilde{g}_{ab}}{F^{\prime}}\right]=\int
d^{4}x\sqrt{|\tilde{g}|}\left[\frac{1}{F^{\prime}}\,\frac{1}{2}\tilde{g}^{ab}\,\partial_{a}\psi\,
\partial_{b}\psi-\frac{1}{(F^{\prime})^{2}}\,\frac{1}{2}m_{\psi}^{2}\psi^{2}\right]\,.\]

The equation of motion for the matter scalar field~$\psi$ follows as
usual from the variational principle $\delta
\tilde{S}_{\text{mat}}/\delta\psi=0$. After some standard
manipulations and partial integration, we arrive at
\begin{eqnarray}\label{eq:variationpsi} \delta{\tilde{S}}_{\text{mat}}= -\int
d^{4}x\sqrt{|\tilde{g}|}\left[\frac{1}{\sqrt{|\tilde{g}|}}
 \partial_{a}\left(\frac{1}{F^{\prime}}\,\sqrt{|\tilde{g}|}\,\tilde{g}^{ab}\,\partial_{b}\psi\,\right)+\frac{m_{\psi}^{2}}
 {(F^{\prime})^{2}}\, \psi\right]\,\delta\psi.\end{eqnarray}
The first term in the bracket at the integrand can be written
as\begin{eqnarray}
 \frac{1}{F^{\prime}}\,\frac{\partial_{a}\left(\sqrt{|\tilde{g}|}\,\tilde{g}^{ab}\,\partial_{b}\psi\right)}{\sqrt{|\tilde{g}|}}
 +\tilde{g}^{ab}\,\partial_{b}\psi\,\partial_{a}\left(\frac{1}{F^{\prime}}\right)
 =  \frac{1}{F^{\prime}}\,\tilde{\square}\psi-\frac{\tilde{g}^{ab}\,\partial_{b}\psi}{(F^{\prime})^{2}}\,\frac{\partial F^{\prime}}
 {\partial\varphi}\,\partial_{a}\varphi\,,\end{eqnarray}
and then using (\ref{eq:defvarphi}) we finally obtain the full
equation of motion for the matter scalar~$\psi$:\[
\tilde{\square}\psi-\tilde{g}^{ab}(\partial_{b}\psi)\sqrt{\frac{2}{3}}\frac{\partial_{a}\varphi}{M_{P}}+\frac{1}{F^{\prime}}\,
m_{\psi}^{2}\psi=0.\]

When the gravitational scalar~$\varphi$ is in a slow-roll regime we
have~$|\partial_{a}\varphi|\lesssim\sqrt{2\,V(\varphi_{s})}=\sqrt{6}\,M_{P}\,r_{e}^{-1}/\sqrt{F^{\prime}}$,
where $V(\varphi_{s})$ was given in~(\ref{eq:Pot-varphi}). The
maximal coupling with matter appears when~$\partial_{a}\varphi$
saturates the slow roll bound, in which case the equation of motion
for $\psi$ in the Einstein frame can be cast in a simplified
notation as follows:
\begin{equation}\label{eq:friction}
\tilde{\square}\psi-2\,\frac{r_{e}^{-1}}{\sqrt{F^{\prime}(\varphi)}}\,(\partial\psi)+\frac{m_{\psi}^{2}}{F^{\prime}(\varphi)}\psi=0\,.\end{equation}
Here, the mass~$m_{\psi,E}=m_{\psi}/\sqrt{F^{\prime}}$ corresponds
to that of the matter field $\psi$ in the Einstein frame and
contains the factor~$\sqrt{F^{\prime}}$. We see from
(\ref{eq:friction}) that this factor is shared by the fifth force
``friction'' coupling of matter ($\psi$) with $\varphi$ in that
frame: $\Gamma_{\psi,E}\sim r_{e}^{-1}/\sqrt{F^{\prime}(\varphi)}$.
Since the dimensionless ratio
$\Gamma_{\psi,E}/m_{\psi,E}=r_{e}^{-1}/m_{\psi}$ is independent
of~$F^{\prime}$, the relative strength of the fifth force acting on
the matter field $\psi$ (measured by comparing it to its mass) is
the same for all frames. Thus, we find
$\Gamma_{\psi,J}=\left(\Gamma_{\psi,E}/m_{\psi,E}\right)\,m_{\psi}=r_{e}^{-1}$
for the effective fifth force strength in the Jordan frame. In other
words, this coupling is essentially given by the present Hubble
rate: $r_{e}^{-1}\approx H_{0}$. Even in the considered case, where
the coupling of $\varphi$ to matter is maximal, we obtained an
extremely weak fifth force. Therefore, we must conclude that such
coupling is completely unobservable.\footnote{One can
show that if we would also transform the matter field $\psi$ into
the Einstein frame, $\psi\to \sqrt{F^{\prime}}\,\tilde{\psi}$, then
the tiny fifth force could be fully absorbed into a very small,
unobservable, correction to the matter field mass.}
It is also interesting to mention the
effect of $\varphi$ on photon interactions with matter. Since the
action for photons is invariant under conformal transformations of
the metric, they are not subject to explicit fifth forces mediated
by $\varphi$. However, they feel the metric deviation in our
modified gravity model with respect to standard General Relativity.
This effect might be detected by gravitational lensing experiments
depending on the strength of the deviation. We briefly address this
possibility in the next section.

\section{Numerical results for the large scale astrophysical domain\label{sec:Numerical-results-betaB}}

The large scale cosmological domain characterized by the FLRW metric
was studied in detail in \cite{BSS-2010,BBS-2011}, and we saw it is dominated
by the $\beta/B$ term in (\ref{eq:Action-Complete}). We have argued
in Sec.~\ref{sec:Spherical-symmetric} that the $\beta/B$ term in the
action is also the dominant term at large length scales for the
metric (\ref{eq:SSS-metric}), which is more in accordance with the
astrophysical environment. Unfortunately, due to the occurrence of
the Gau\ss-Bonnet invariant in (\ref{BRG}) it is difficult to find
exact solutions on a spherically symmetric background.  This is
quite obvious from the complicated structure of the field equations
(\ref{eq:fieldequations})-(\ref{eq:Eabfull}), especially when the
curvature invariants are non-constant. We will show here some
numerical solutions of these equations in the vacuum case and we briefly discuss the deviation from the standard SdS solution. The latter metric, as given by equations (\ref{eq:SSS-metric}) and~(\ref{eq:SdS-Func-a}),
will be used for setting the initial conditions for large
radius~$r$, i.e.\ the situation for which the $\beta$-terms dominate
over the $\beta_{\odot}$ ones. This procedure ensures a smooth
transition to the cosmological de~Sitter result.
\begin{figure}[th!]
\begin{centering}
\hspace*{3.5mm}\includegraphics[clip,width=0.87\textwidth]{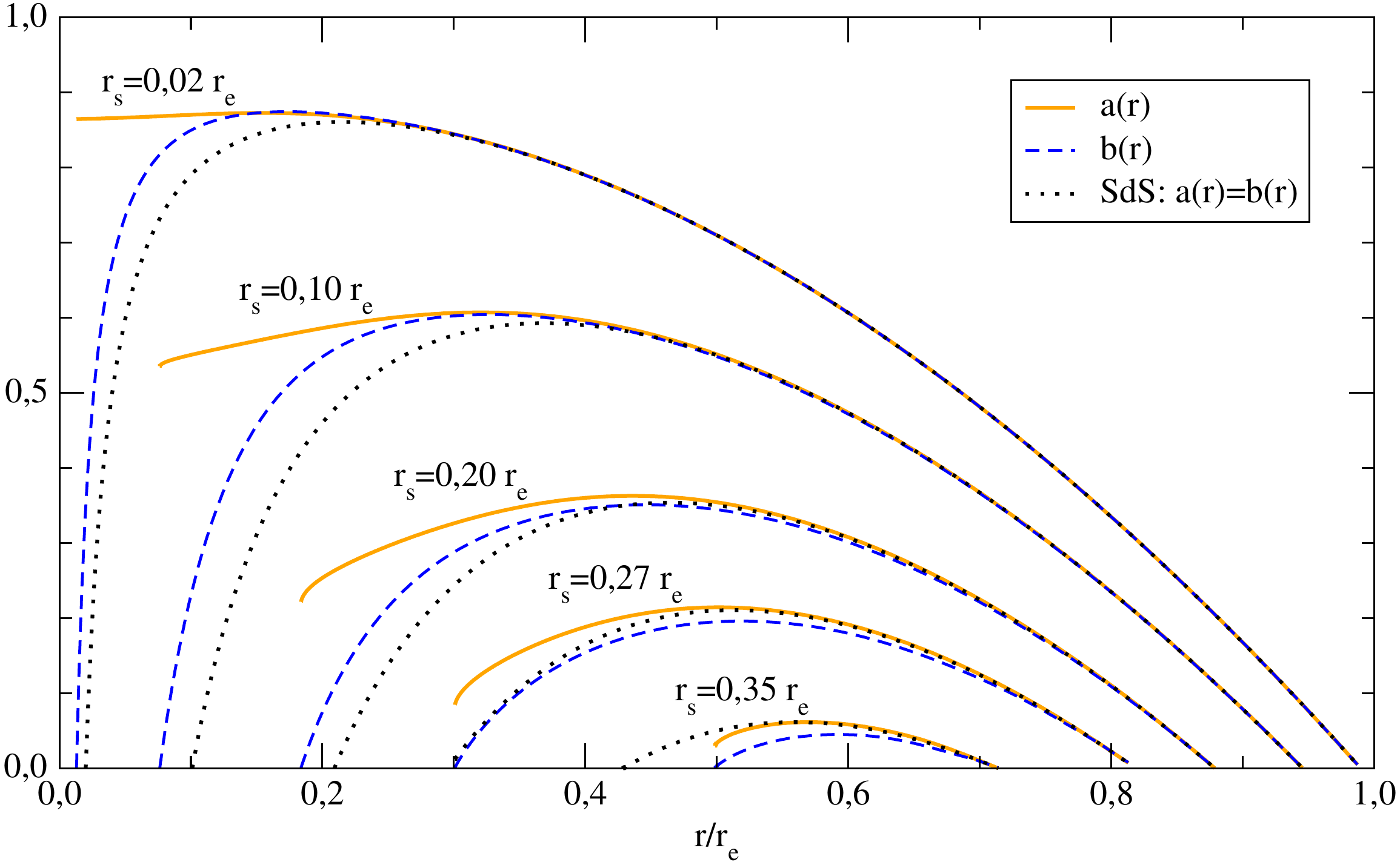}\\
\vspace{0.25cm}
\includegraphics[clip,width=0.9\textwidth]{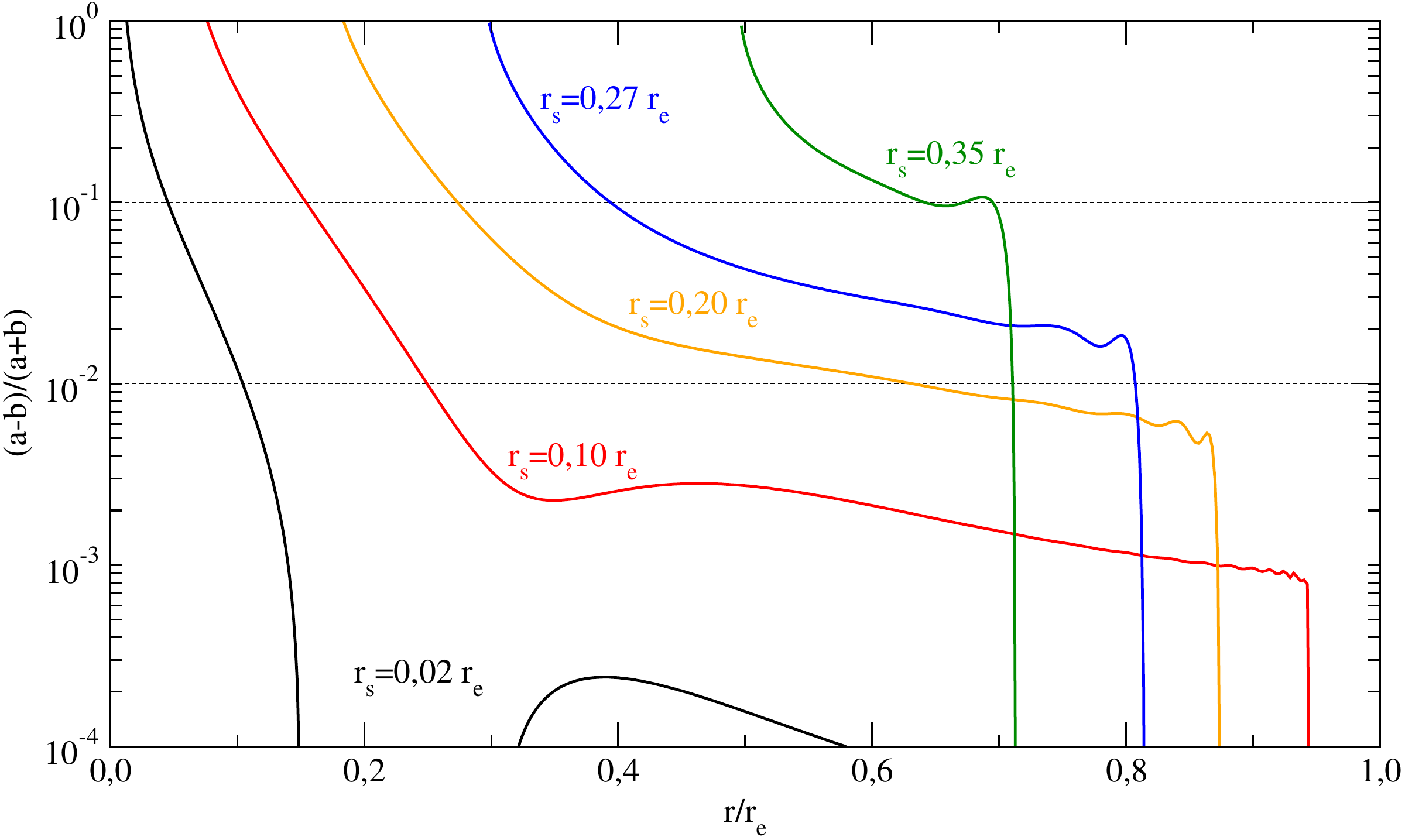}
\par\end{centering}
\caption{{\bf Top}: numerical solutions for the functions $a(r)$ and
$b(r)$ in the spherically symmetric metric~(\ref{eq:SSS-metric}) for
the pure $\beta/B$ model without matter. The initial conditions have
been fixed close to the higher zero of the corresponding SdS
solution~(\ref{eq:SdS-Func-a}), which is plotted, too.  {\bf
Bottom}: relative differences $\frac{a(r)-b(r)}{a(r)+b(r)}$ of the
numerical functions $a(r)$ and $b(r)$. Note that the numerical error
dominates around the higher zeros of both
functions.\label{fig:Numerical-a-b}}
\end{figure}

Specifically, to obtain the pure effects of the~$\beta/B$ term, we
set~$\beta_{\odot}$ to zero in (\ref{eq:Eabfull}), and the initial
conditions for the metric are set very close but below the higher
zero of the function $a(r)$ in Eq.~(\ref{eq:SdS-Func-a}). For small
values of the Schwarzschild radius parameter~$r_{s}\ll r_{e}$ this
is approximately at~$r=r_{e}$. As we know already, $r_{s}=0$
corresponds to the de~Sitter solution, which is exact also for
the~$\beta/B$ model. Therefore, we consider only solutions with
non-zero~$r_{s}$. Some examples are shown in
Fig.~\ref{fig:Numerical-a-b}, where the two functions $a(r)$ and
$b(r)$ in the metric~(\ref{eq:SSS-metric}) are displayed together
with the standard SdS result. Let us note that in the case of small
perturbations, one usually writes $a(r)=1+2\Psi(r)$ and
$b(r)=1+2\,\Phi(r)$ where $\Psi(r)$ and $\Phi(r)$ are the two
gravitational potentials associated to the metric. While in the
standard SdS case they are equal, in the present case we look for
deviations $a(r)-b(r)=2[\Psi(r)-\Phi(r)]$ that could provide
information of the modified gravity model. As expected, these
deviations grow with increasing~$r_{s}$. While~$b(r)$ follows
qualitatively the standard result, the function~$a(r)$ stays at
larger values especially for small~$r$, which we identify as an
effect of the $r$-dependence in~$B$. The relative difference between
both metric functions is plotted also in
Fig.~\ref{fig:Numerical-a-b} (bottom), it is of order one around the
lower zero of~$b(r)$. However, we expect that this singular point
must lie within the matter distribution, whose metric has to be
matched to the vacuum solution in Fig.~\ref{fig:Numerical-a-b} at a
larger radius~$r$. Whether this matching is possible and how the
inner solution for the metric behaves, depends strongly on the
properties of the matter distribution. We keep this question open
for a future investigation. \newtext{At the moment} we cannot exclude the possibility that at intermediate (intergalactic) scales some non-trivial corrections to standard expectations might creep in. We provide some discussion of this possibility in the next section.

\section{Possible gravity modifications at large distances\label{sec:GravityModif}}

Let us return for a moment to the discussions made in Sect. \ref{sec:Spherical-symmetric}. We have seen that while both at local astrophysical scales and at very large cosmological domains the relaxation mechanism works perfectly well and can be studied analytically, at intermediate scales our knowledge is more limited and we need to resort to the numerical analysis within some approximations, as we have seen in the previous section. The characteristic distance entering the potential deviations is
\begin{equation}\label{eq:rcritical}
r_{c}\sim\left(r_{s}^{2}r_{e}^{2}\frac{\beta_{\odot}}{\beta}\right)^{1/6}\,,
\end{equation}
where we know that $\beta$ and $\beta_{\odot}$ are fixed in order of magnitude and (same) sign by the condition that the relaxation mechanism works in the cosmological and astrophysical domains, respectively. These conditions entail the relations $|\beta|\sim \rLi\,H_0^4$ and $|\beta_{\odot}|\sim \rLi\,H_0^2$. Hence it follows from (\ref{eq:rcritical}) that
\begin{equation}\label{eq:rcritical2}
r_{c}\sim\left(\frac{r_{s}^{2}r_{e}^{2}}{H_0^2}\right)^{1/6}\sim\left(\frac{r_s}{H_0^2}\right)^{1/3}=\left(\frac{G_N\,M_s}{H_0^2}\right)^{1/3}\,,
\end{equation}
where $M_s$ is the mass of the galaxy or cluster of galaxies sourcing the gravitational field. We may compare this length scale with the one at which one expects deviations to the Newtonian gravity at large distances in ordinary extended gravity theories, i.e.\ with a late-time modification of the gravitational interaction\,\cite{ModifiedGrav1-01,ModifiedGrav1-02,ModifiedGrav1-03,ModifiedGrav1-04,ModifiedGrav2-01,ModifiedGrav2-02,ModifiedGrav2-03,ModifiedGrav2-04}. These effects have been specifically addressed in Ref.\,~\cite{Navarro:2005gh} by considering a Lagrangian modification of the form $\delta{\cal L}=M_P^2\,\mu^{4n+2}/Q^n$ with $Q=R^{abcd}R_{abcd}$ the square Riemann tensor. For $n=1$ this is not far away from the behavior of the gravity modification in our relaxation model (\ref{eq:Action-Complete}) at large distances, except that in our case we have both a gravity modification and the large CC in mutual interplay. It is however illustrative to compare the two kind of models for $n=1$\,\footnote{The CC relaxation model has also been analyzed at the cosmological level in more general cases in which we have an arbitrary power of the denominator (\ref{BRG}) and also when a power of the Ricci scalar $R$ is included in the numerator of the gravity modification in (\ref{eq:Action-Complete}), see \cite{BSS-2010,BBS-2011} for details. However, here we just analyzed the canonical, i.e.\ the simplest realization of these models at the astrophysical level, and therefore we should naturally compare also with the simplest case $n=1$ of Ref.~\cite{Navarro:2005gh}. }, for which the the characteristic scale of the induced modification of Newtonian gravity is
\begin{equation}\label{eq:rcriticalNavarro}
r_{c}^{*}\sim\left(\frac{\left(G_N\,M_s\right)^{2}}{H_0^{3}}\right)^{1/5}\,,
\end{equation}
after setting $\mu\sim H_0$, which is the extremely small value that is necessary to choose for the $\mu$-parameter in order to explain the late-time cosmic acceleration in this kind of models ~\cite{Navarro:2005gh}. The ratio between the two characteristic length scales is
\begin{equation}\label{eq:ratioNavarroRelax}
\frac{r_{c}^{*}}{r_c}\sim\left(G_NM_s\,H_0\right)^{1/15}={\cal O}(0.1)\,.
\end{equation}
For the numerical evaluation of this ratio we have used the source mass of a relatively large galaxy (say, $M_s\sim10^{11}\,M_{\odot}$, i.e.\ with a number of solar masses similar to our galaxy), or the mass of a large cluster of similar galaxies. In both cases the two length scales do not differ in more than one order of magnitude, so they are indeed very close. In other words, in both classes of models (relaxation model and ordinary late-time extended gravity) we expect deviations from Newtonian gravity at similar large scales in the $100$ kpc-Mpc range, as it follows from the above formulae. Let us remark, however, that in the relaxation model we have a double bonus which is absent in the ordinary case, to wit: i) we do not need to use extremely small scales as $\mu\sim H_0\sim 10^{-33}$ eV but rather scales ${\cal M}\sim$ meV and ${\cal M}'\sim$ MeV both lying in the natural SM range (cf. Sect. \ref{sec:Spherical-symmetric}); and ii) the local astrophysical domain is protected from unwanted vacuum effects even in the presence of the large $\rLi$. Indeed, the typical value of $r_c$ emerging from (\ref{eq:rcritical2}) is roughly from $10$ to $100$ kpc in the case of a galaxy, whereas for a large cluster it may reach the few Mpc level. As these length scales are at the same time the characteristic sizes of a galaxy and a cluster of galaxies, respectively, it follows that the local astrophysical domains remain well protected from large vacuum effects in our relaxation models. Finally, in the intermediate region beyond these distances (say up to $\sim 10-100$ Mpc) we cannot study the problem in a simple way, as the numerical analysis of the previous section has shown,
because the SdS metric is no longer a good approximate solution. Only after attaining very large distances (of order of a few hundred Mpc
at least) we retrieve again the relaxation mechanism in the
cosmological domain\,\cite{BSS-2010} with all its phenomenological success\,\cite{BBS-2011}.

\section{Conclusions}

In this letter, we have performed \newtext{a first} investigation of the modified gravity implementation of the CC relaxation mechanism in a static and spherically symmetric background, which serves as an approximation
for the study of the mechanism at length scales characteristic of
the astrophysical domain. We found the mechanism to be working in
the sense that the space-time curvature is not dominated by the
initial large vacuum energy density~$\rho_{\Lambda}^{i}$. Instead, a
small effective dark energy density
$\mathcal{O}(M_{P}^{2}r_{e}^{-2})\sim M_P^2\,H_0^2\sim\rLo$ is
found, similar to our earlier results in a cosmological
background\,\cite{BSS-2010}. Moreover, we studied the influence of
fifth forces on massive objects, which turned out to be long ranged,
but extremely weak and therefore inconspicuous. This shows that the idea
of CC relaxation may be useful in more general contexts than just in
cosmology. Given the unconventional starting point, one could not
naively expect this local astrophysical behavior from the beginning.
Interestingly, we also found that at very large scales of order of
$10^{-2}\,H_0^{-1}$ the metric may deviate significantly from the
standard SdS solution, thus opening up a possible way to detect this
mechanism via gravitational lensing \newtext{and other effects}.

Possible observable implications at these scales have already been proposed in other modified gravity theories~\cite{Navarro:2005gh}. \newtext{In the light of these studies} one can foresee deviations from the standard Newtonian/GR behavior at length scales of $100$~kpc-Mpc. We cannot exclude, for instance, the possible implications of these deviations e.g.\ by changing the required abundances of dark matter in the intergalactic domains in order to explain the DM problem in the large. However, the investigation of these potentially relevant issues goes far beyond the content of the present Letter. Their more precise description is therefore an important future challenge in the CC relaxation approach.

Finally, some theoretical issues remain open for further study; in particular, the fact that the relaxation mechanism exerts such a dominating influence. In this sense, further insight should go into the direction of finding ways to complete this model and understanding the role of the vacuum energy versus the matter sources. Quite in contrast to the present approach, the usual modifications of gravity disclaim \textit{ab initio} any understanding of how to cope with the huge vacuum energy injected in the universe by QFT/string theory, and just focus on late-time simulations of the measured effect. \newtext{This way is therefore no more satisfactory}. A link between the two points of view is still missing, and hence more work is required to understand the ultimate interplay between gravitation, vacuum energy and matter in our cosmos. We hope that our approach may offer a new perspective for an eventual solution of this
difficult problem.

\vspace{0.3cm}

{\bf Acknowledgments} \vspace{0.2cm}

FB and JS would like to thank D. Polarski for very useful
discussions. The authors have been partially supported by DIUE/CUR
Generalitat de Catalunya under project 2009SGR502; FB and JS also by
MEC and FEDER under project FPA2007-66665 and by the
Consolider-Ingenio 2010 program CPAN CSD2007-00042, and HS also by
the Ministry of Education, Science and Sports of the Republic of
Croatia under contract No. 098-0982930-2864.

\newcommand{\JHEP}[3]{ {JHEP} {#1} (#2)  {#3}}
\newcommand{\NPB}[3]{{\sl Nucl. Phys. } {\bf B#1} (#2)  {#3}}
\newcommand{\NPPS}[3]{{\sl Nucl. Phys. Proc. Supp. } {\bf #1} (#2)  {#3}}
\newcommand{\PRD}[3]{{\sl Phys. Rev. } {\bf D#1} (#2)   {#3}}
\newcommand{\PLB}[3]{{\sl Phys. Lett. } {\bf B#1} (#2)  {#3}}
\newcommand{\EPJ}[3]{{\sl Eur. Phys. J } {\bf C#1} (#2)  {#3}}
\newcommand{\PR}[3]{{\sl Phys. Rep. } {\bf #1} (#2)  {#3}}
\newcommand{\RMP}[3]{{\sl Rev. Mod. Phys. } {\bf #1} (#2)  {#3}}
\newcommand{\IJMP}[3]{{\sl Int. J. of Mod. Phys. } {\bf #1} (#2)  {#3}}
\newcommand{\PRL}[3]{{\sl Phys. Rev. Lett. } {\bf #1} (#2) {#3}}
\newcommand{\ZFP}[3]{{\sl Zeitsch. f. Physik } {\bf C#1} (#2)  {#3}}
\newcommand{\MPLA}[3]{{\sl Mod. Phys. Lett. } {\bf A#1} (#2) {#3}}
\newcommand{\CQG}[3]{{\sl Class. Quant. Grav. } {\bf #1} (#2) {#3}}
\newcommand{\JCAP}[3]{{ JCAP} {\bf#1} (#2)  {#3}}
\newcommand{\APJ}[3]{{\sl Astrophys. J. } {\bf #1} (#2)  {#3}}
\newcommand{\AMJ}[3]{{\sl Astronom. J. } {\bf #1} (#2)  {#3}}
\newcommand{\APP}[3]{{\sl Astropart. Phys. } {\bf #1} (#2)  {#3}}
\newcommand{\AAP}[3]{{\sl Astron. Astrophys. } {\bf #1} (#2)  {#3}}
\newcommand{\MNRAS}[3]{{\sl Mon. Not. Roy. Astron. Soc.} {\bf #1} (#2)  {#3}}
\newcommand{\JPA}[3]{{\sl J. Phys. A: Math. Theor.} {\bf #1} (#2)  {#3}}
\newcommand{\ProgS}[3]{{\sl Prog. Theor. Phys. Supp.} {\bf #1} (#2)  {#3}}
\newcommand{\APJS}[3]{{\sl Astrophys. J. Supl.} {\bf #1} (#2)  {#3}}

\newcommand{\Prog}[3]{{\sl Prog. Theor. Phys.} {\bf #1}  (#2) {#3}}
\newcommand{\IJMPA}[3]{{\sl Int. J. of Mod. Phys. A} {\bf #1}  {(#2)} {#3}}
\newcommand{\IJMPD}[3]{{\sl Int. J. of Mod. Phys. D} {\bf #1}  {(#2)} {#3}}
\newcommand{\GRG}[3]{{\sl Gen. Rel. Grav.} {\bf #1}  {(#2)} {#3}}

\end{document}